\documentclass[]{raa}
\usepackage{graphicx,times}
\usepackage{natbib}

\providecommand{\bm}[1]{\mbox{\boldmath$#1$\unboldmath}}
\def\kms{$\rm{km~s}^{-1}$}

\makeatletter
\newcommand{\figcaption}{\def\@captype{figure}\caption}
\newcommand{\tabcaption}{\def\@captype{table}\caption}
\makeatother

\begin{document}

\title{The Inversion of the Real Kinematic Properties of Coronal Mass
Ejections by Forward Modeling}

\volnopage{ {\bf 2010} Vol.\ {\bf 10} No. {\bf XX}, 000--000}
   \setcounter{page}{1}
   \author{Wu, Y.
      \inst{1}
   \and Chen, P. F.
      \inst{1,2}
   }
\institute{Department of Astronomy, Nanjing University, Nanjing 210093,
	China; {\it chenpf@nju.edu.cn}\\
        \and
     Key Lab of Modern Astron. and Astrophys., Ministry of Education,
		China\\
\vs \no
  {\small Received [2010] [month] [day]; accepted [year] [month] [day]}
}

\abstract{The kinematic properties of coronal mass ejections (CMEs)
suffer from the projection effects, and it is expected that the real
velocity should be larger and the real angular width should be smaller
than the apparent values. Several attempts have been tried to correct
the projection effects, which however led to a too large average
velocity probably due to the biased choice of the CME events. In order
to estimate the overall influence of the projection effects on the
kinematic properties of the CMEs, we perform a forward modeling of the
real distributions of the CME properties, such as the velocity, the
angular width, and the latitude, by requiring their projected
distributions to best match the observations. Such a matching is
conducted by Monte Carlo simulations. According to the derived real
distributions, it is found that (1) the average
real velocity of all non-full-halo CMEs is about 514 \kms, and the
average real angular width is about 33$^\circ$, in contrast to the
corresponding apparent values of 418 \kms and 42.7$^\circ$ in
observations; (2) For the CMEs with the angular width in the range of
$20^\circ-120^\circ$, the average real velocity is 510 \kms and the
average real angular width is 43.4$^\circ$, in contrast to the
corresponding apparent values of 392 \kms and 52$^\circ$ in
observations.
}
\authorrunning{Wu, Y. \& Chen, P. F.}
\titlerunning{The Inversion of the Real Kinematic Properties of CMEs}
\maketitle

\section{Introduction}
\label{sect:Introduction}

Expulsions of dense clouds of plasma from the solar corona, which are
termed coronal mass ejections (CMEs), have been explored in the past
four decades. Most CMEs originate in the lower corona, and expel about
$10^{14}-10^{16}$ grams of plasma with the kinetic energy about
$10^{30}$ to $10^{32}$ ergs \citep{Huds+06}, and the velocity
ranging from tens to more than three thousand \kms \citep{St.C+00}.
The CME phenomenon has provoked great interests within the 
communities of solar and solar-terrestrial physics since CMEs are often
associated with other activities, such as flares, filament eruptions,
radio bursts, and solar energetic particle events
\citep[e.g.,][]{Munr+79, kahl92, Chen+06, Gopa+08, Chen+09}. Whether a
CME can approach the Earth and induce a strong geomagnetic storm depends
on various properties of the CME, e.g., the location of the source
region, the angular width, mass, velocity, magnetic field intensity, and
so on. 

Since the first detection on 1971 December 14 by the seventh
Orbiting Solar Observatory (OSO-7) coronagraph \citep{Tous+73},
CMEs have been observed by several other space instruments \citep{Yash+08}.
The Large angle and Spectrometric Coronagraph (LASCO) on
board the Solar and Heliospheric Observatory (SOHO) mission has made a
significant impact on the understanding of CMEs since its first light in
1996 January. The total number of CMEs that LASCO observed is over 14
thousand until the end of 2009 February. The huge database, e.g., the
CME catalog in NASA/CDAW\footnote{http://cdaw.gsfc.nasa.gov/CME\_list},
is very useful for investigating various properties of CMEs. For 
example, \citet{Yash+04} found that the average velocity of normal
CMEs (with the angular widths between $20^\circ-120^\circ$) is about
428 \kms, and the corresponding average angular width is $56^\circ$ for
the CME events observed during 1996-2002. Although the statistical
results might change more or less as more observations are made or
different methods are applied in identifying the CME events, research on
the CME kinematic properties based on this catalog should have a high
degree of reliability. However, the measurements of the CME kinematic
properties directly from coronagraph observations generally suffer from
the projection effects. For example, it would be expected that the real
CME velocity could be higher and the real angular width could be lower
than the apparent values \citep{Vrsn+07}.

One straightforward method to get the real kinematic properties of CMEs
is to study the limb CME events, which are nearly free of the
projection effects. \citet{Burk+04} identify a sample of 111
``limb" events among 1351 CMEs observed by the Solar Maximum Mission
(SMM) in 1980 and 1984-1989. The limb events have an average velocity of
519 \kms and an average angular width of $52.3^\circ$, in contrast to 349
\kms and $46^\circ$ for all CMEs observed by SMM. It is understandable
that the velocity of the limb events is larger than that of all events,
however, the result of the angular width is not consistent with the
projection effects, which might be due to the limb sample is biased to
the eruptive filament association and therefore is not representative
\citep{Burk+04}. A similar survey with a larger sample is strongly in
need. It is noted that, however, uncertainty may arise in identifying
the limb events in this kind of survey. If EUV dimmings are used to
identify the source region of the CMEs, many slow events without
evident dimmings would be missed.

Several attempts have been tried to correct the projection effects for
individual events. For halo events, \citet{Mich+03} presented a
correction method and found that halo CMEs have an average real velocity
of 1080 \kms and average real angular width of $120^\circ$, in contrast
to 900 \kms and $360^\circ$ before the correction. For non-halo CMEs,
\citet{Lebl+01} proposed a correction method, where the locus of
the associated flare is used to determine the center of the source
region of each CME. \citet{Yeh+05} slightly modified such a method
and applied it to 619 CME/flare events where the loci of the flares are
available in the Solar Geophysical Data Report. They found that the
average velocity of the sampled CMEs is 792 \kms, and the average angular
width is $59^\circ$, compared to 549 \kms and $77^\circ$ before
correction. It is noted that their surveys may be biased to strong
events (so that the loci of the flares were well documented), which
results in a remarkably high average velocity. 

One common requirement for the aforementioned inversion methods is that
one needs to determine the central position of the CME source region,
which is not easy to be determined for a large sample. In order to
correct the projection effects without the knowledge of the central
position of the source region for each CME event, we propose a forward
modeling technique in this paper, and apply it to derive the real
velocity and angular width. This paper is divided into four parts.
The forward modeling method is described in Section \ref{sect:Method},
and the results are shown in Section \ref{sect:Results}, which is
followed by discussions in Section \ref{sect:Discussions}.

\section{Method}
\label{sect:Method}

The directly measurable quantities of CMEs are the apparent velocity
($v$), the apparent angular width (AW), and the measurement position
angle (MPA).  With a large sample of events, we can get their
distributions statistically, which are fitted with respective formulae.
Our forward modeling method can be described as follows: Assuming that
the real velocity, real angular width, and the real latitude of sample
CMEs follow the same functions as their apparent counterparts, but with
different parameters, we project the sample CMEs into the plane of the
sky to get the projected distributions of their kinematic quantities.
The parameters are adjusted by trial and error until the projected
distributions of $v$, AW, and MPA best fit the observed distributions.

In order to get the statistical results of the observations, we appeal
for the NASA/CDAW CME catalog, which is based on SOHO/LASCO
observations. There are 14120 CMEs from 1996 January 11 to 2009 February
28 in this catalog. Among these events, 106 CMEs, whose velocity can not
be detected, and 396 full halo CMEs are excluded, by which we obtain a
sample of 13618 CMEs. The distributions of their apparent velocity,
apparent angular width, and MPA are showed in Figure \ref{Fig1}, which
can be fit with lognormal, Gaussian, and double Gaussian functions,
respectively, as shown in Equations (\ref{eq1})-(\ref{eq3}). The average
apparent velocity of the whole sample is about 418 \kms, and the average
apparent angular width is about $42.7^\circ$.

\begin{equation}
f(v)={\frac{1}{0.57\sqrt{2\pi}v}}\exp{[-\frac{(\ln{v}-5.85)^2}{2\times 0.57^2}]}
\label{eq1}
\end{equation}
\begin{equation}
f(AW)=\frac{1}{60.7\sqrt{2\pi}}\exp[-\frac{AW^2}{2\times 60.7^2}]
\label{eq2}
\end{equation}
\begin{equation}
f(MPA)=\frac{1}{30.5\sqrt{2\pi}}\exp[-\frac{(MPA-90)^{2}}{2\times 30.5^{2}}]+\frac{1}{30.5\sqrt{2\pi}}\exp[-\frac{(MPA-270)^{2}}{2\times 30.5^{2}}]+0.003
\label{eq3}
\end{equation}

   \begin{center}
   \includegraphics[width=14cm, angle=0]{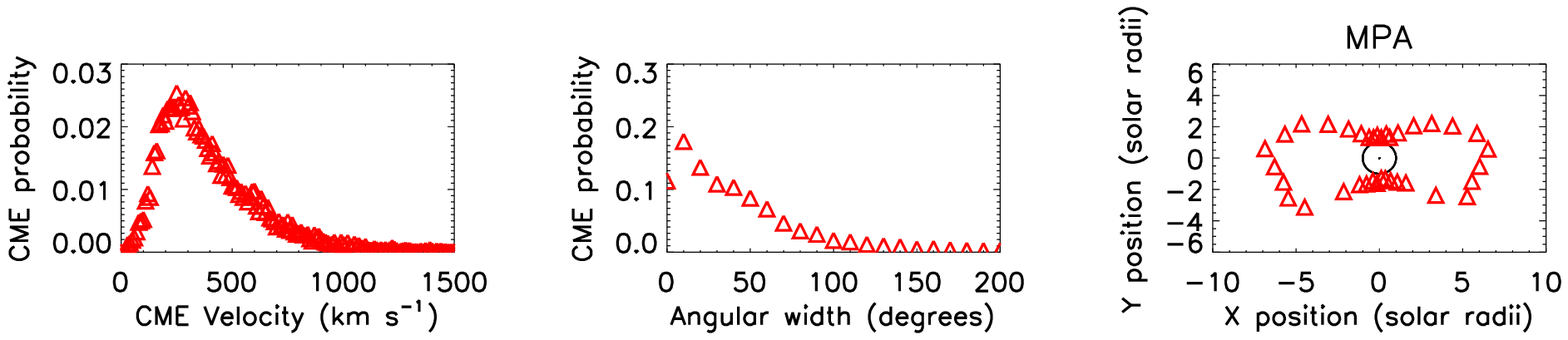}
  \figcaption{Distributions of the apparent velocity ({\it left panel}),
    the apparent angular width ({\it middle panel}), and the MPA ({\it 
    right panel}) of the sample CMEs. The velocity is binned in 10 \kms
    intervals, the angular width is in $10^\circ$ intervals, and the MPA
    is in $10^\circ$ intervals. In the right panel, the distance of each
    triangle from the solar disk center represents the fraction of CMEs
    in this direction, and the circle marks the solar limb.}
   \label{Fig1}
   \end{center}
 
Since most CMEs propagate radially with an almost fixed angular width,
the so-called cone model is often applied to describe the geometry of
CMEs \citep{Howa+82, Fish+84, Mich+03, Burk+04, Schw+05, Vrsn+07, Gao+09}. With the assumption of the cone model, the
kinematic properties of a CME in 3-dimensional space are characterized
roughly by 4 variables, i.e., the real velocity ($v_{re}$), the real 
angular width ($AW_{re}$), the latitude ($\theta$) and the longitude 
($\phi$) of the source center. It is reasonable to assume that CME
sources are randomly distributed along the longitude in the
heliocentric coordinates. For other three variables, we assume that the
distributions of $v_{re}$, $AW_{re}$, and $\theta$ take the same
functions as those of the apparent velocity, angular width, and the MPA
(Equations \ref{eq1}-\ref{eq3}), but with modified parameters, which
are expressed as follows:

\begin{equation}
F(v_{re})=\frac{1}{v_{re}\sigma_1\sqrt{2\pi}}\exp[-\frac{(\ln{v_{re}}
	-\mu)^2}{2\sigma_1^2}],
\label{eq4}
\end{equation}

\begin{equation}
F(AW_{re})=\frac{1}{\sigma_2\sqrt{2\pi}}\exp(-\frac{AW_{re}^2}
	{2\sigma_2^2}),
\label{eq5}
\end{equation}

\begin{equation}
F(\theta)=\frac{1}{\sigma_{3}\sqrt{2\pi}}\exp(-\frac{\theta^{2}}{2\sigma_{3}^{2}}),
\label{eq6}
\end{equation}
\noindent
where $\mu$, $\sigma_1$, $\sigma_2$, and $\sigma_3$ are parameters to
be determined.

In order to quantitatively calculate the projection effects on the CME
velocity and angular width, we perform Monte Carlo simulations, which
are described as follows: One million artificial CMEs are created, whose
real velocity, angular width, and latitude of the central source region
satisfy the distributions presented in Equations (\ref{eq4}-\ref{eq6}),
and whose longitude of the central source region is randomly 
distributed in the range of $0^\circ$-$360^\circ$. Each of the
artificial CMEs is then projected into the plane of the sky, through
which the distributions of apparent velocity, apparent angular width,
and the MPA can be derived. The four parameters ($\mu$, $\sigma_1$, 
$\sigma_2$, and $\sigma_3$) are adjusted until the projected
distributions reproduce the observational counterparts (Equations
\ref{eq1}-\ref{eq3}), where the similarity is examined with the 
Kolmogorov test.

It is noted that, when projecting the artificial CMEs into the plane of
the sky, the relation between the apparent velocity and the real value
depends on the choice of the cone models \citep{Schw+05,Vrsn+07,Gao+09}.
We assume that CMEs are symmetric, and propagate out radially through
the corona, and their leading edge is a part of sphere, connecting
the cone tangentially like an ice-cream, i.e., the Model C in 
\citet{Schw+05}. Besides, according to \citet{Zhan+10}, some CMEs,
especially those near the Sun-Earth line, would be too faint to be
detected by coronagraphs due to Thomson scattering, which should be
taken into account in our study. For that purpose, we calculate the
white-light intensity of each CME leading edge at $3R_\odot$ in the
plane of the sky. Those CMEs whose white-light intensity falls below the
$3\sigma$ of the fluctuation level of the solar wind are considered to
be missed by coronagraphs \citep[see][for details]{Zhan+10}. The
expression for calculating the Thomson-scattered intensity is given
by \citet{Bill66}. In our work we neglect the limb darkening effect,
so the expression can be simplified as follows,

\begin{equation}
I=\pi\sigma J_{0}n_{0}R_{0}\frac{1}{81}\int^{\pi/2}_{0}(2C\cos^{3}\theta
	-A\cos^{5}\theta),
\label{eq:thom}
\end{equation}
\noindent
where $\sigma$ represents the Thomson scattering cross section, $J_{0}$
the photosphere radiative intensity at solar surface, $n_{0}$ the
electron density, $R_{0}$ the solar radius, $\theta$ the angle between
the plane of the sky and the direction of CME propagation, $A=\frac{
\cos^{2}\theta\sqrt{9-\cos^{2}\theta}}{27}$, and $C=\frac{4}{3}-\frac{
\sqrt{9-\cos^{2}\theta}}{3}-\frac{(9-\cos^{2}\theta)^{3/2}}{81}$.

\section{Results}
\label{sect:Results}

The derived real distributions of the CME velocity, the angular width,
and the latitude are expressed as follows:

\begin{equation}
F(v_{re})=\frac{1}{0.53\sqrt{2\pi}v}\exp[-\frac{(\ln{v_{re}}-6.09)^2}
	{2\times 0.53^2}],
\label{eq:realv}
\end{equation}

\begin{equation}
F(AW_{re})=\frac{1}{37.2\sqrt{2\pi}}\exp(-\frac{AW_{re}^2}{2\times 
	37.2^2}),
\label{eq:realaw}
\end{equation}

\begin{equation}
F(\theta)=\frac{1}{26.0\sqrt{2\pi}}\exp(-\frac{\theta^2}{2\times
	26.0^2}).
\label{eq:reallat}
\end{equation}

Their profiles are depicted in the left column of Figure \ref{Fig2}. As
these quantities are projected into the plane of the sky, the resulting
distributions of the apparent velocity, the angular width, and the MPA
are shown in the right column of Figure \ref{Fig2} as crosses, where
they are compared to the LASCO observations ({\it triangles}). It is
seen that the simulated apparent distributions of the CME velocity, the
angular width, and the MPA match the observations very well, which 
implies that the derived distributions in Equations 
(\ref{eq:realv}--\ref{eq:reallat}) might well represent the real 
situation.

   \begin{center}
   \includegraphics[width=14cm, angle=0]{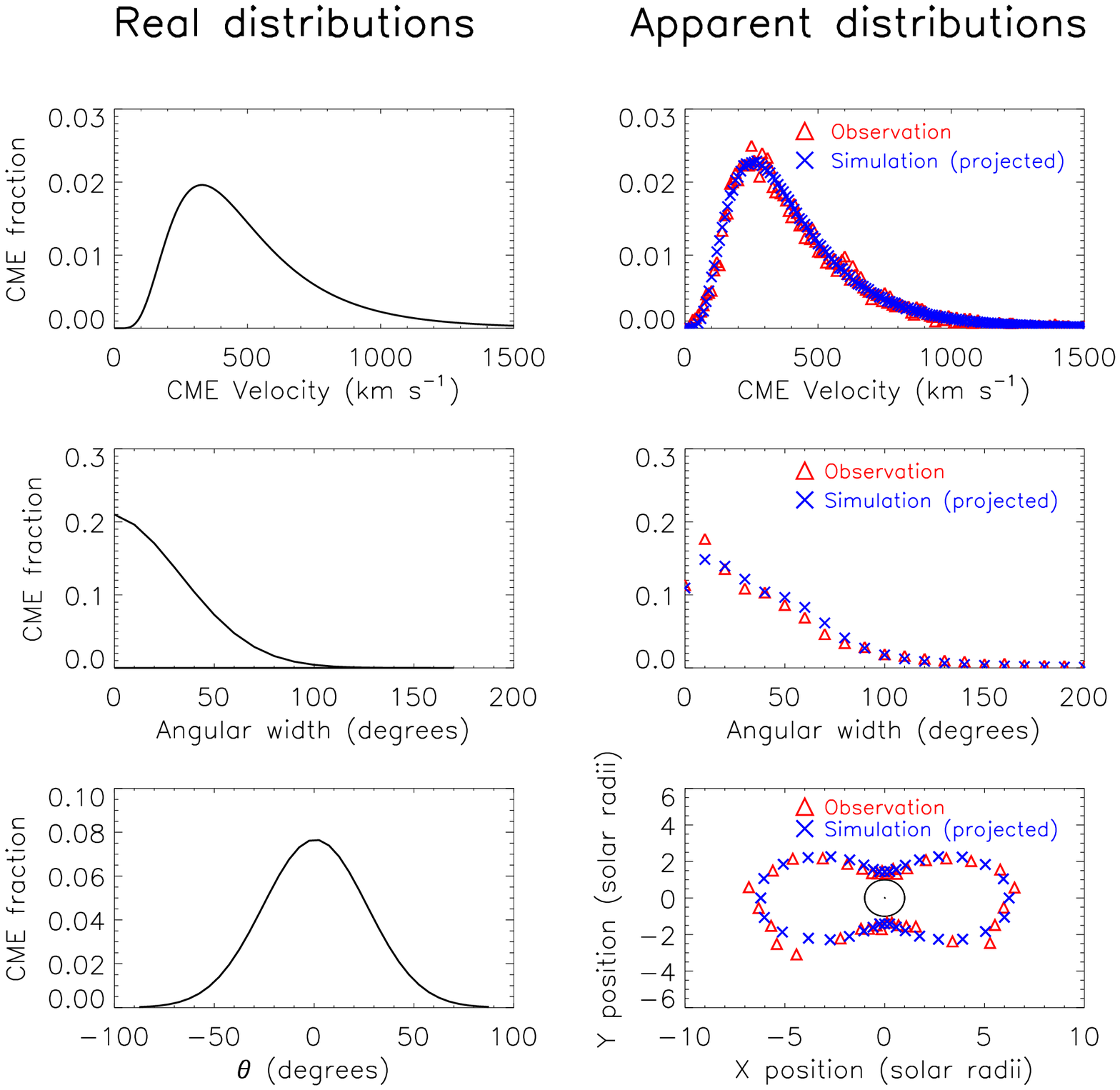}
   \figcaption{{\it left column}: The derived real distributions of
	the CME velocity ({\it top}), the angular width ({\it middle}),
	and the latitude $\theta$ ({\it bottom}); {\it Right column}:
	The apparent distributions of the CME velocity ({\it top}), the
	angular width ({\it middle}), and the MPA ({\it bottom}), where
	the triangles correspond to the SOHO/LASCO observations, and the 
	crosses to the projection of the real distributions.}
   \label{Fig2}
   \end{center}

According to the derived real distributions of the CME velocity and the
angular width ($AW_{re}$) for all non-full-halo CMEs, the average CME
velocity is 514 \kms, which is $\sim 23\%$ larger than the apparent
value, and the average angular width is $33^\circ$, which is $\sim 23
\%$ smaller than the apparent value. In particular, for those CMEs whose
angular widths are between $20^\circ$ and $120^\circ$ \citep[the
so-called normal CMEs in][]{Yash+04}, the corrected average CME velocity
is 510 \kms, which is $\sim 30\%$ larger than the apparent value (392
\kms), and the corrected average angular width is $43.4^\circ$, which is
$\sim 17\%$ smaller than the apparent value ($52^\circ$).

It is noticed that the real angular width distribution is peaked near
$0^\circ$, however, its apparent angular width is peaked around 
$10^\circ$, which is consistent with the observations. Our simulations
reveal that the significant drop of the number of narrow CMEs (with the
angular width less than $10^\circ$) is due to that some of them are two
weak to be detected when they propagate not far from the Sun-Earth
direction.

\section{Discussions}
\label{sect:Discussions}

The accumulating observations of CMEs enable the statistics of their
properties, such as their angular size and their kinematics. However,
the measurements of the CME geometry and kinematics suffer from the
projection effects, and it is expected that the real angular width
should be smaller, whereas the real velocity should be larger. In this
paper, with a forward modeling by Monte Carlo simulations, we corrected
the distributions of the CME velocity and the angular width for the
projection effects, as revealed by Equations
(\ref{eq:realv}--\ref{eq:reallat}). It is found that (1) for all
non-full-halo CMEs, the
real velocity is averaged at 514 \kms, and the real angular width is
averaged at $33^\circ$, in contrast to 418 \kms and 42.7$^\circ$ before
correction; (2) for CMEs with the apparent angular width between
$20^\circ-120^\circ$, the real velocity is averaged at 510 \kms, and the
real angular width is averaged at $43.4^\circ$, in contrast to 392 \kms
and 52$^\circ$ before correction. The average real velocity of CMEs
derived in this paper is similar to that of the limb events, which are
free of the projection effects \citep{Burk+04}.

It should be noted that in our forward modeling, it was assumed for
simplicity that the three prescribed variables (i.e., the CME velocity,
the angular width, and the latitude of the central source region) are
completely independent. With a large sample, \citet{Yeh+05} did find
that there is little correlation between the CME velocity and the 
angular width. However, \citet{Yash+04} and \citet{Vrsn+07} claimed
that wider CMEs tend to be faster. In order to clarify the discrepancy,
we collect the 13618 non-full-halo CME events observed by SOHO/LASCO
during the period 1996 January - 2009 February and display the scatter
plot of the relation between the observed CME angular width and the
velocity in the left panel of Figure \ref{Fig3}. It is found that the
two quantities present a very weak positive correlation, with the
correlation coefficient being only 0.23. We also divide the sample into
a subsample with CMEs observed near solar maximum and another subsample
with CMEs near solar minimum, it is also revealed that the CME velocity
and the angular width present little correlation. For comparison, the
right panel of Figure \ref{Fig3} displays the scatter plot of the CME
velocities and the angular widths of the artificial CMEs in our Monte
Carlo simulations. It can be seen that the distributions of the scatter
points in the two panels of Figure \ref{Fig3} are quite similar. The
linear Pearson correlation test indicates that the similarity between
the two panels is as high as 0.95. It is noted that the sparse data
points with $v>1000$ \kms or $AW>120^\circ$ are not shown in Figure
\ref{Fig3} in order to make the figure clear, though they are included
in our study.

   \begin{center}
   \includegraphics[width=14cm, angle=0]{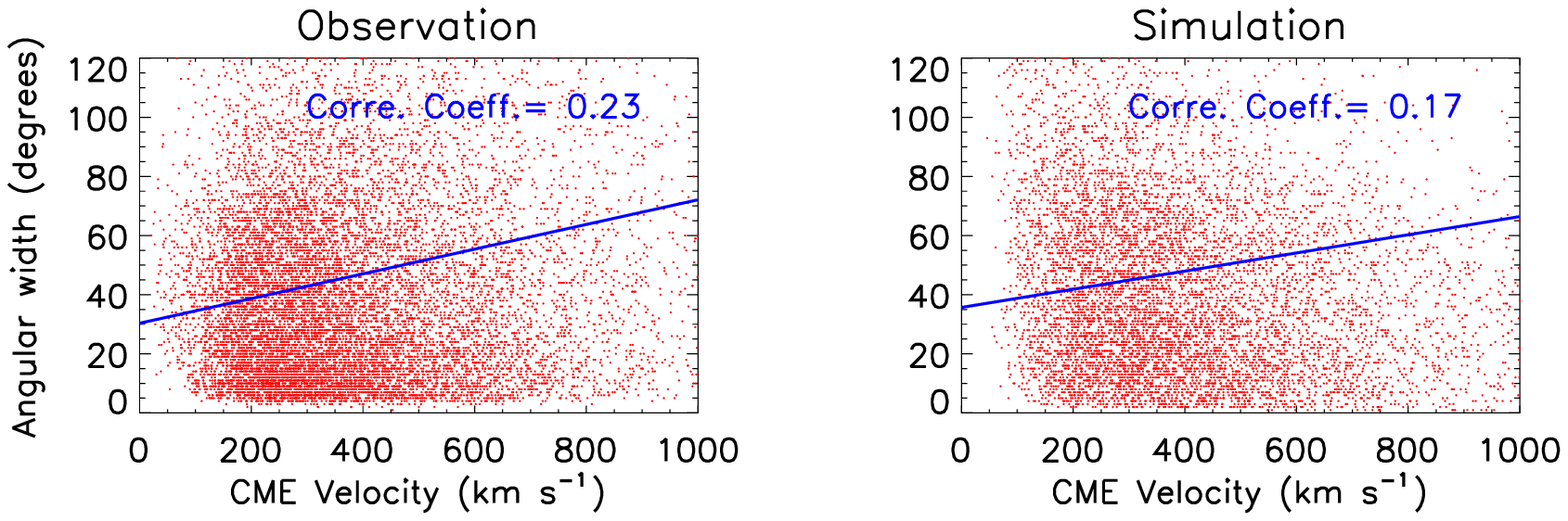}
   \figcaption{Scatter plots of the relation between the CME velocity
	and the angular width in observation ({\it left}) and
	simulation ({\it right}). The solid lines represent the linearly
	fitted lines.}
   \label{Fig3}
   \end{center}

In Figure \ref{Fig4} we compare the correlations between the CME
velocity and the MPA in both observations and simulations. The left
panel displays the scatter plot in the observations, where we can see
that the correlation coefficient between the CME velocity and the MPA
is almost zero. The right panel shows the corresponding scatter plot in
our Monte Carlo simulations. The linear Pearson correlation test
indicates that the similarity between the distributions in the two
panels is as high as 0.91. Similarly to Figure \ref{Fig3}, the sparse
data points with $v>1000$ \kms or $MPA>200^\circ$ are not shown in 
Figure \ref{Fig4}, though they are included in our study.

   \begin{center}
   \includegraphics[width=14cm, angle=0]{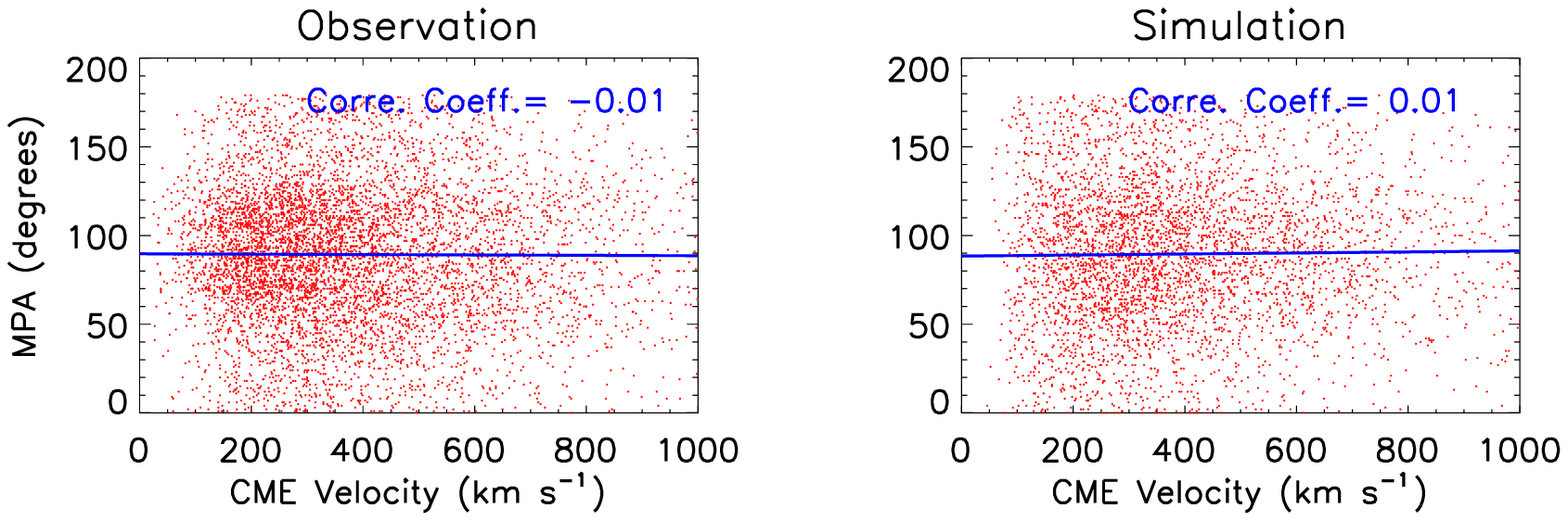}
   \figcaption{Scatter plots of the relation between the CME velocity
	and the MPA in observation ({\it left}) and simulation ({\it
	right}). The solid lines represent the linearly fitted lines.}
   \label{Fig4}
   \end{center}

It is noted in passing that the CME angular width and the MPA also show
little correlation. Therefore, with the largest-ever sample that
includes almost all the CMEs observed by the SOHO/LASCO coronagraph, we
found that the CME velocity, the CME angular width, and the latitude of
the central source region are all mutually uncorrelated. Such a lack of
correlation validates our assumptions used in our forward modeling in
deriving the CME real velocity and angular width distributions.

\begin{acknowledgements}
We are grateful to C. Fang and M. D. Ding for discussions and
suggestions throughout the work. This research is supported by the
Chinese foundations 2006CB806302 and NSFC (10403003, 10933003, and
10673004). The CME catalog used in this paper is generated and
maintained at the CDAW Data Center by NASA and The Catholic University
of America in cooperation with the Naval Research Laboratory. SOHO is a
project of international cooperation between ESA and NASA. 
\end{acknowledgements}

\label{lastpage}
\end{document}